# ON FACEBOOK, MOST TIES ARE WEAK

PASQUALE DE MEO
Department of Ancient and Modern Civilizations, University of Messina, Italy
EMILIO FERRARA
School of Informatics and Computing, Indiana University, Bloomington, USA
GIACOMO FIUMARA and ALESSANDRO PROVETTI
Department of Mathematics and Informatics, University of Messina, Italy
E-mail addresses: ferrarae@indiana.edu, {pdemeo,gfiumara,ale}@unime.it

*The emergence of pervasive socio-technical networks brings new conceptual and technological challenges. A central research theme is the evaluation of the intensity of relations that bind users and how these facilitate communication and the spread of information. These aspects have been extensively studied in social sciences before, under the framework of the strength of weak ties theory proposed by Mark Granovetter. Some authors recently investigated whether Granovetter's theory can be extended to online social networks like Facebook, suggesting using interaction data to predict tie strength. Those approaches require handling user-generated data that are often not publicly available for privacy reasons. We propose an alternative definition of weak and strong ties that only requires knowledge of the topology of the social network, e.g., who is friend with whom on Facebook. Our approach relies on the fact that OSNs tend to fragment themselves into communities. Hence, we suggest classifying as weak ties those edges linking individuals belonging to different communities, while strong ties are those connecting users in the same community. We tested this definition on a large network representing part of the Facebook social graph. We studied how weak and strong ties affect the information diffusion process. Our findings suggest that individuals in OSNs self-organize to create well-connected communities, while weak ties bring about cohesion and optimize the coverage of information spread.*

The analysis and understanding of Online Social Networks (OSNs) such as Facebook finds a theoretical foundation in *Social Network Analysis* [15]. However, studying a real OSN poses several *computer science* challenges, given the size, distribution and organization (privacy, visibility rules, etc.) of the data available to the regular OSN subscriber [1]. In such a context, the analysis of large subsets of an OSN should generate a series of statistically robust measurements that form the basis of the understanding of OSNs structure and evolution. Indeed, the obtained aggregate measures shall be very valuable in Data management, Privacy management and Online Marketing. A challenging problem is the evaluation of the intensity of relations that bind users and how these facilitate the spread of information. These aspects have been extensively studied in social sciences before, notably with the *strength of weak ties* theory proposed by Mark Granovetter [13].

Weak ties are connections between individuals who belong to distant areas of the social graph, i.e., which happen to have most of their relationships in different national, linguistic, age or common-experience groups. Weak ties are powerful tools to transfer information across large social distances and to wide segments of the population. Vice versa, strong ties are contacts between trusted/known persons (e.g., family ties or close friendships). We ask whether Granovetter's weak ties are also to be found in OSNs like Facebook in the form he envisioned them, i.e., connections between individuals who belong to different areas of the social graph. Such a question, however, is hard to answer for at least two reasons. First, Facebook is mainly organized around the recording of just one type of relationship: *friendship*. This implies that Facebook friendship captures (and compresses) several degrees and nuances of human relationships that are hard to separate and characterize through an analysis of online data. Second, as Facebook is growing in size and complexity, its friendship network is growing denser, not sparser [1].

As OSNs become more and more interconnected, testing Granovetter's theory poses serious scalability challenges. Early research works [9] used a supervised approach, where a panel of Facebook users was asked to assess the strength of their own friendship ties. Large-scale studies of Granovetter's theory in the fashion of [9] would arguably be very hard to conduct, given the sheer size of today's OSNs. Other approaches, notably [2], which accessed to Facebook own data on user activities and computed the tie strength as a function of type and frequency of user interactions. However, a cut-off threshold is required to distinguish strong ties from weak ones and the tuning of that threshold has a crucial impact on the correct identification of weak ties.

In this article, we propose a new definition of weak ties that is rooted in the analysis of large OSNs and aware of the computational challenges lying thereof. The starting point is that in both online and off-line social networks participants tend to organize themselves into *dense communities* [8]. We propose to first identify communities within the network and second to classify as *weak ties* those edges that connect users located in different communities; *strong ties* will be those edges between users in the same community. We argue that through our definition, identifying weak ties becomes *i)* fast, thanks to the efficiency of recent algorithms for finding communities in large networks [8] and *ii)* robust, because no threshold needs to be defined. We performed extensive experimental analysis, thanks to a public dataset on Facebook friendship released by [10] and a null-model comparison against randomly generated graphs. Two well-known community detection algorithms, namely the *Louvain Method* (LM) [3] and *Infomap* [20], were deployed. From the analysis of the experimental results, we report the following findings:

1. The weak (resp., strong) ties discovered by LM tend to coincide with those found by Infomap. Importantly, our definition of weak ties is thus largely independent of the particular choice of the community-detection algorithm adopted.
2. Overall, our community-defined weak ties outnumber the strong ties.
3. Weak ties occur more frequently in communities of small size.
4. Weak ties identified by our approach play a crucial role to spread information over a network and their removal reduces the portion of the network that can be reached by information diffusion.

## Weak and Strong Ties

Mark Granovetter gave the now-classic definition of *strength of a social tie* [13]:

*The strength of a tie is a (probably linear) combination of the amount of time, the emotional intensity, the intimacy (mutual confiding), and the reciprocal services that characterize the tie.*

This definition introduces some important features of a social tie like the intensity of the connection (i.e., the frequency of contacts) and its reciprocity. Granovetter himself considered the identification of strong and weak ties by means of topological information related to the social network structure. To that purpose, he introduced the concept of *bridge* as follows:

*A bridge is a line in a network which provides the only path between two points. Since, in general, each person has a great many contacts, a bridge between A and B provides the only route along which information or influence can flow from any contact of A to any contact of B.*

Notice that, according to Granovetter's definition, *all bridges are weak ties*. With OSNs, unfortunately, Granovetter's definition of bridge is restrictive and unsuitable. In fact, the well-known *small world effect* (i.e., the presence of short-paths connecting any pair of vertices) and the *scale-free degree distribution* (i.e., the presence of *hubs* that maintain the whole network effectively connected) make it unlikely to find an edge whose deletion would completely disconnect its terminal vertices.

To adapt Granovetter's definition to OSNs, we could re-define a *shortcut bridge* as a link that connects a pair of vertices whose deletion would simply cause an increase of the distance (defined as the length of the shortest path linking two vertices) between them. This sensible definition is, in our opinion, more controversial than it seems. First, it depends on the notion of shortest path but, unfortunately, the identification of all-pairs shortest paths is computationally unfeasible even on networks of modest size. Second, even the alternative definition of distance, i.e., that of *weighted path*, (computed as the sum of the weights assigned to each connection that is used) would remain computationally prohibitive for OSNs. Our goal is to explore a novel and computationally feasible definition of weak ties that suits the analysis of large-scale OSNs. Instead of a *relational* definition, which is based on the intensity of the interactions between two users, we propose a *community-based* definition. We define *weak ties* as those edges that, after dividing up the network into communities (thus obtaining the so-called *community structure*), connect vertices belonging to different communities. Finally, we classify intra-community edges as *strong ties*.

One of the most important features of our weak ties is that those that are bridges (in Granovetter's sense) create more, and shorter, paths: their deletion would be more disruptive than the removal of a strong tie, from a community-structure perspective. In retrospect, this may be the reason why weak ties (albeit defined in a slightly different fashion) have been recently proved to be very effective in the diffusion of information [5,21].

### *Benefits of our definition*

Our definition of weak tie has four appealing features:
– it is weaker than Granovetter's, viz. the fact that vertices linked by a weak tie belong to different communities does not imply that the edge between them is a bridge: actually, its deletion may not disconnect its vertices (it almost never does, in practice).

– it enables the weak/strong classification on the basis of topological information *only*. [21] also used topological information but only *locally*, i.e., the neighbors of the two terminal vertices, whereas our definition handles *global information*, as it relies on the partitioning of the whole network.

- it is *binary* because it labels each edge in the network as either weak or strong. As a consequence, we do not need to ``tune'' any threshold, below which edges are classified as weak (the upshot being that we cannot compare two edges on the basis of their strength).

- it depends on the accuracy of the community discovery phase, for which accurate and scalable algorithms are now available [8]. Our experiments show that our definition of weak tie is robust with respect to the choice of the particular community detection algorithm.

## Related Approaches

In the literature, several works have examined the strength of weak ties in terms of non-topological information. A non-topological approach requires to choose some *measurable variables* by which the strength of the relation binding two users can be deduced. In such scenario, weak ties are intended in a relational sense, i.e., they connect acquaintances who do not frequently interact, and, therefore, are not strongly influencing each another. For "off-line" social networks, [17] identified some variables such as communication reciprocity and the presence of at least one mutual friend as indicators of a weak tie.

In the Social Web context [19] and [9] studied small OSNs (56 and 35 participants, respectively) and assigned weights to the relationships on the basis of several measures of strength, e.g., intimacy/closeness, reciprocity, sociability/conviviality, etc. that were assessed through direct questions to the participants. In particular, [9] extended Marsden-Campbell's method to Facebook by identifying as many as 74 variables as potential predictors of strength. Then, they modeled strength as a linear combination of the above-mentioned variables and weights were computed by means of a variant of Ordinary Least-Square regression. To validate their model, [9] recruited 35 users who were asked to *rate* their Facebook friends. They achieved an accuracy of about 85% but their performance seem hard to replicate on large OSN fragments. Crucially, their method requires the collection of a large number of detailed information on users' behaviors; today, due to privacy reasons and to the limitations on the usage of proprietary data, most of the required data are unlikely be available to academic studies.

[21] formalized the strength of ties with a weight $w_{ij}$ assigned to the edge connecting vertices *i* and *j* as follows:

$$w_{ij} = \frac{c_{ij}}{k_i + k_j - 2 - c_{ij}}$$

where $k_i$ and $k_j$ are the degrees of *i* and *j* and $c_{ij}$ is the number of their mutual acquaintances. Numerical simulations showed that by gradually deleting ties with the lower *w* values, information coverage dropped sharply.

Their results corroborate the hypothesis that weak ties are in fact the key to information diffusion. A recent work on weak ties in Facebook due to [2] considered 4 parameters to measure the strength of a connection: *(i)* frequency of private messages, *(ii)* frequency of public comments left on each other's posts, *(iii)* number of times they jointly appear in a photo and *(iv)* number of times they jointly commented a third-party post.

Unlike us, [2] do not consider, at least explicitly, topological aspects and their analysis requires access to proprietary data and records of users' activity; thus, we cannot easily compare our methodologies. Our work is similar to [21] and [2] only in the fact that *i)* weak ties are understood as useful connectors that favor the spread of information and *ii)* both our work and [2] are experimentally tested on Facebook.

Vice versa, a relevant difference emerges: both authors assign scores to ties, and classify them according to a threshold value. In contrast, we classify ties as weak or strong depending on whether they connect vertices located in different communities or not; our classification scheme is *binary* and it does not use scores and threshold, which may be hard to set up and tune properly. Another recent approach in the literature is due to Grabowicz *et al.* [12].

Similarly to us, they used information about the network topology to identify weak ties. However, their study focuses on Twitter, which can be modeled as a directed network where user relationships are mostly asymmetric. Peculiar Twitter features, i.e. *re-tweet* and *mention* have a major impact on the identification of weak ties; therefore, our study and [12] are not easily comparable.

## Results

We present the results of the experimental tests that we carried out to highlight the pros and cons of our definition of weak tie. For the preliminary community detection phase we deployed two popular algorithms, the Louvain method (LM) [3] and Infomap [20]. Two test-beds were considered and compared against: i) a fragment of the Facebook network collected by [10] and consisting of 957,000 thousands users and 58.4 millions friendship connections[1] and *ii) null-model* networks consisting of Erdos-Renyi random graphs with a number of vertices up to 2048; the probability of having an edge between an arbitrary pair of vertices was varyied uniformly from 0.05 to 0.95 with a 0.05 step.

---

[1] For a complete description of the dataset please see the Supplementary Material.

*Robustness of Weak Ties Definition*

As a preliminary experiment, we studied the *robustness* of our definition of weak ties in respect to the method adopted for finding communities.

Different community detection methods might produce different results, albeit slightly so, hence our weak/strong ties classification could become dependent on the community-detection method. We ran both LM and Infomap on our Facebook sample: the community structures found by the two algorithms were compared applying the Normalized Mutual Information (NMI). From our experiments we found an NMI around 0.9, which informs us that the level of disagreement between the two algorithms is quite low[2].

As a result, we can conclude that on Facebook our definition of weak ties is *robust* because it does not depend on the choice of the community-detection algorithm.

However, we have to note that the communities discovered by LM and Infomap might differ in subtle ways and still be characterized by a high NMI.

Therefore, next, we will present a detailed, graphical comparison of results we achieved by applying LM and Infomap, so as to better understand what are the practical consequences arising from the adoption of a particular community detection algorithm.

*Distribution of strong and weak ties in Facebook*

Here we study the distribution of weak and strong ties in Facebook. To this purpose, we computed the *complementary cumulative distribution function* (CCDF) of weak (resp., strong) ties, i.e., the probability of finding a vertex that has more than $k$ weak (resp., strong) ties. The obtained results are plotted in Figures 1a and 1b.

The CCDF associated with strong ties decreases faster than that for weak ones. At $k = 4$ we see a tipping point after which weak ties quickly outnumber strong ties, i.e., the latter are much less frequent in higher-degree vertices.

The results of our experiments agree fairly well with Granovetter's vision of weak ties. In fact, sociological theories such as Triadic Closure [13], Cognitive Balance [14], and Homophily [18] suggest that individuals tend to aggregate in small communities.

According to these theories, we can explain that the intensity of human relations is very tight within small groups while it decreases towards individuals belonging to distant communities. Therefore, most connections are weak in Granovetter's sense: small amount of contacts, low frequency of interactions, etc.

*Weak and Strong Ties distribution in Random Graphs*

We here present a comparative analysis of weak/strong ties distribution in Erdos-Renyi's random graphs, i.e., we check whether the results obtained on Facebook still holds on graphs whose structure is known in advance and lacks, by construction, a clear community structure. We computed the ratio $R_{avg}$ of the number of weak ties to the total number of ties and Figures 2a and 2b report $R_{avg}$ for different values of $|V|$ and when the probability $p_{link}$ of having an edge between an arbitrary pair of vertices uniformly varies from 0.05 to 0.95.

We can conclude the following:
1. $R_{avg}$ is always greater than 0.6, i.e., weak ties still outnumber strong ones also in case of randomly generated graphs.
2. $R_{avg}$ is relatively stable and independent of $p_{link}$, i.e., the sparsity of $G$ has a limited impact on the number of weak ties.
3. $|V|$ has a limited impact on $R_{avg}$: in fact, when $|V|$ goes from 128 to 4096 (i.e., it increases by a factor of 32), $R_{avg}$ only increases by 17.14%.

*The Role of Weak Ties in Information Diffusion*

In the next experiment, we studied how weak ties influence the information diffusion process. This study clarifies the connection between our definition and Granovetter's, since weak ties are deemed to provide specific links between individuals who would otherwise remain disconnected as they belong to distant areas of the social graph.

Granovetter's weak ties should play a central role in the spread of information, but what about our weak ties? How well do they convey information?

To assess their role, we applied the *Independent Cascade Model* (ICM) [11] to simulate the information propagation over a network, again taking the Facebook dataset and Erdos-Renyi's random graphs for comparison. In ICM, a vertex $v_0$ is selected uniformly at random to forward a message to its neighbors, with probability equal to $p_{inf}$ (infection probability).

---
[2] The definition of NMI is reported in the Supplementary Material; NMI ranges between 0 and 1 and if it approaches 1 then the communities found by the two algorithms can be regarded as coinciding.

Each "infected" vertex can, in turn, recursively propagate the message to its neighbors. According to [16], reasonable values of $p_{inf}$ are 0.01, 0.02 and 0.03.

To generate statistically significant results, we selected the vertex $v_0$ to start from 250 times; at each selection of $v_0$, we simulated the propagation of a message. We measured the *coverage* $\sigma$ defined as the ratio of the number of vertices receiving a message (infected vertices) to the total number of vertices. The experiment was repeated by progressively (and randomly) deleting a fraction $\tau$ of weak ties.

In our simulation, $\tau$ ranged from 0.1 to 0.9 and, for each value we computed the corresponding coverage. The whole procedure was repeated separately for weak and strong ties. In Figures 3a and SM1 we report the values of $\sigma$ obtained by applying the LM and Infomap on Erdos-Renyi's random graphs, respectively. Due to space limitations, we consider only a graph with $|V| = 512$ vertices, $p_{link} = 0.05$ and $p_{linf}=0.03$.

Notice that the gradual deletion of weak ties yields a decrease of $\sigma$: on average, the largest decrease of $\sigma$ 11.98% with an average decrease equal to 5.71% (with standard deviation equal to 3.4%). The same behavior emerges when removing strong ties, but the decrease is less marked. We observed that if $|V|$ increases, the coverage $\sigma$ also increases.

The values of $\sigma$ associated with our Facebook dataset are reported in Figures 3b and SM2.

Once again, for a fixed value of $\tau$, the removal of weak ties yields a decrease of $\sigma$ more marked than the removal of strong ones. The largest observed decrease of $\sigma$ was about 14.31%, with an average decrease equal to 6.26% (standard deviation is equal to 3.1%).

These values are always greater and those for Erdos-Renyi's random graphs. This proves that the removal of weak ties is more significant in real OSNs rather than in networks not equipped with a meaningful community structure. In the light of this experiment, we may conclude that our definition of weak ties captures Granovetter's idea: deleting weak ties decreases/obstructs the flow of information much more than removing strong ones.

## Conclusions

In this article we presented a novel definition of weak ties designed for OSNs like Facebook that is based on the community structure of the network itself. Our experiments carried out on a large Facebook sample and on randomly generated graphs highlighted the role and importance of weak ties.

In particular, we characterized the overall statistical distribution of weak ties, as a function of the size of the communities and their density.

We studied their role in information diffusion processes; the results suggest a connection between our definition of weak ties for OSNs and Granovetter's original intuition.

Even though several recent works have focused on the Facebook social graph [4,1], its community structure [7], and also on weak ties *per se* [2], we believe that our community-based definition of weak ties better fits Facebook and similarly large (and dense) OSNs. As for future works, two projects may follow up the results reported here. The first development concerns investigating the applicability of network-weighting strategies, so that the strength of ties can be computed according to a given rationale, for example the ability of each link to spread information. To do so, we intend to adopt a novel method of weighting edges suited for OSNs we devised in [6]. The second interesting development concerns the analysis of the geographical data related to Facebook users.

Thanks to the merging of different graphs, e.g., *social* and *geographical*, we can get additional insights on the role of physical vs. virtual distances.

## Acknowledgments

Thanks to Minas Gjoka for making the Facebook dataset available and to the anonymous reviewers for their careful and constructive comments.

**Figures**

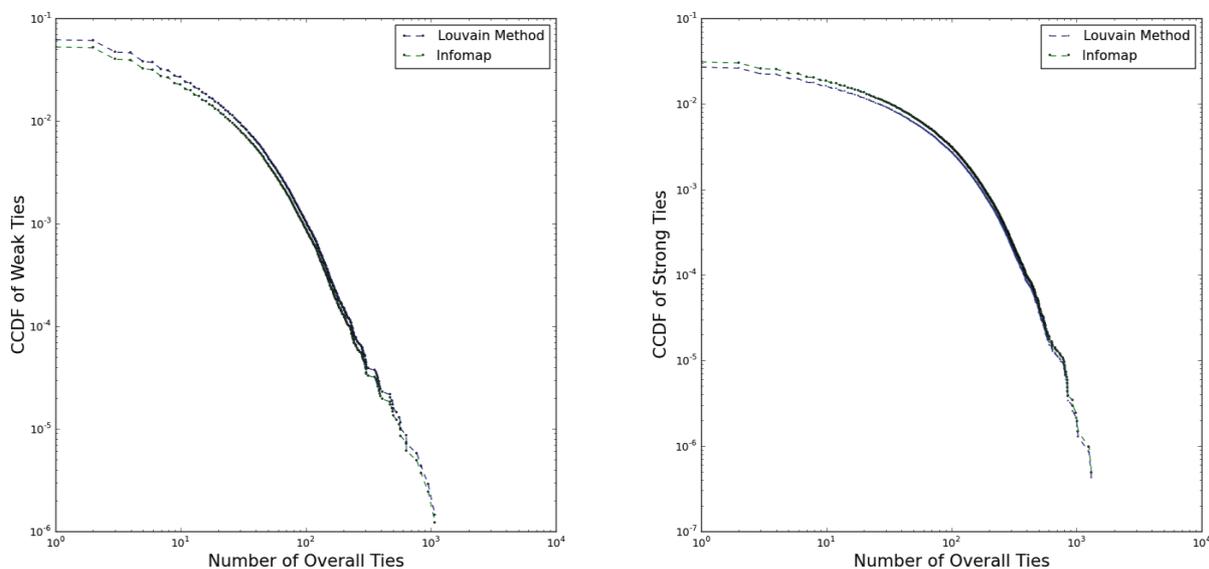

**Fig 1**: The Complementary Cumulative Distribution Function associated with (a) weak ties and (b) strong ties in our Facebook dataset.

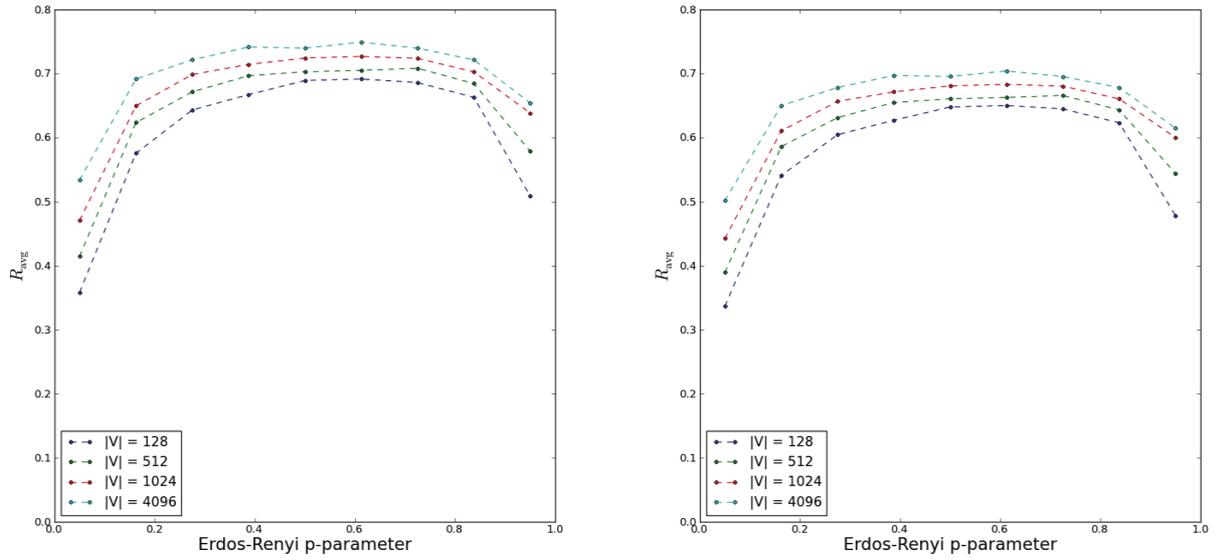

**Fig 2**: The $R_{avg}$ ratio for Erdos-Renyi random graphs clustered with (a) LM and (b) Infomap; instances with $|V|$ = 128, 512, 1024 and 4096.

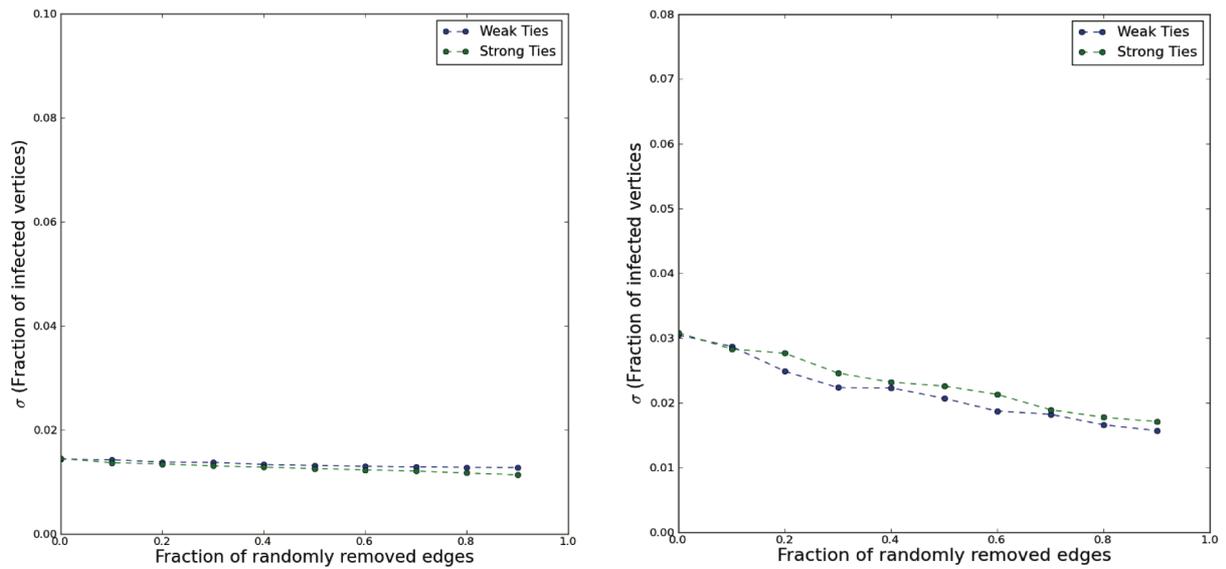

**Fig 3**: (a) Coverage associated with Erdos-Renyi's random graphs ($|V|$ = 512, $p_{link}$ = 0.05, $p_{inf}$ = 0.01) when τ ranges from 0 to 0.9. (b) Coverage associated with the Facebook dataset generated with τ ranging from 0 to 0.9. Diagrams (a) and (b) report the results for the LM.